\documentclass[referee]{raa}

\usepackage{natbib}
\usepackage{color}
\usepackage{amssymb} 
\usepackage{graphicx,times}
\usepackage{epstopdf}
\usepackage{epsfig}
\usepackage{graphicx,color}

\begin{document}

\title{Propagating Disturbances along fan-like coronal loops in an active region}

\volnopage{Vol.0 (200x) No.0, 000--000}
\setcounter{page}{1}

\author{S.~Mandal
    \inst{1}
        \and T.~Samanta
    \inst{1}
        \and D.~Banerjee
    \inst{1}
        \and S. Krishna Prasad
    \inst{2}
    \and L. Teriaca
    \inst{3}
          }
\institute{Indian Institute Of Astrophysics, Bangalore-560 034, India. e-mail:\textcolor{blue}{sudip@iiap.res.in}\\
           \and Astrophysics Research Centre, School of Mathematics and Physics, Queen's University Belfast, Belfast BT7 1NN, UK\\
           \and Max-Planck-Institut fuer Sonnensystemforschung (MPS)  Justus-von-Liebig-Weg 3,  D-37077 Goettingen}

\date{Received~~XXXXXXX; accepted~~XXXXXXX}

\abstract{Propagating disturbances are often observed in active region fan-like coronal loops.
They were thought to be due to slow mode MHD waves based on some of the observed properties. 
But the recent studies involving spectroscopy indicate that they could be due to high speed quasi-periodic upflows which are difficult to distinguish from upward propagating slow waves. In this context, we have studied a fan loop structure in the active region AR 11465 using simultaneous spectroscopic and imaging observations from  Extreme-ultraviolet Imaging Spectrometer (EIS) on board Hinode and Atmospheric Imaging Assembly (AIA) on board SDO. Analysis of the data shows significant oscillations at different locations. We explore the variations in different line parameters to determine whether the waves or flows could cause these oscillations to improve the current understanding on the nature of these disturbances.
}
\maketitle

\section{Introduction}
Coronal loops are made up of hot plasma controlled by magnetic fields. According to magnetohydrodynamic (MHD) wave theory these loops can support different MHD wave modes \citep{2000SoPh..193..139R,2007SoPh..246....3B}. With the recent high resolution observations, quasi-periodic propagating disturbances (PDs) are commonly observed in the solar atmosphere \citep{1998ApJ...501L.217D,2012A&A...546A..93G,2011A&A...528L...4K,2012SoPh..281...67K}. The apparent propagation speed of these PDs ranges from 50 to 200 km~s$^{-1}$ which is close to the sound speed in the corona  \citep{2012SoPh..279..427K}. This led to their interpretation as slow magneto-acoustic waves. Time series analysis show that these quasi-periodic PDs in coronal loops have periods ranging from 3 to 30 min \citep{2001A&A...380L..39B,2009ApJ...696.1448W,2009A&A...503L..25W,2009A&A...493..251G,2012A&A...546A..50K}. Spectroscopic observations reveal that these longitudinal disturbances often  show a correlation between the line intensity and 
Doppler shift \citep{2010ApJ...721..744K}. It was suggested that these oscillations are due to the leakage of the p-mode oscillations which are modified in the presence of magnetic filed and travel to higher atmosphere along the loops \citep{2003ApJ...599..626B,2004Natur.430..536D,2010MNRAS.405.2317S}. The signatures of damping in these PDs have also been reported \citep{2003A&A...408..755D,2002ApJ...574L.101W,2014ApJ...789..118K}. 

Recent spectroscopic observations indicate that these PDs not only show periodic oscillations in intensity and Doppler velocity but sometimes also in line width \citep{2011ApJ...727L..37T}. Although in the past, oscillations observed in the loops were generally interpreted as signatures of various modes of MHD waves,  \citet{2011ApJ...727L..37T} suggested that the observed quasi-periodic oscillations are not necessarily due to the slow magnetoacoustic waves. They reported that the footpoint regions of the loops show a coherent behaviour in all the four line parameters (line intensity, Doppler shift, line width and profile asymmetry) based on which they proposed that PDs can also be due to high-speed quasi-periodic upflows. They found that there are some faint enhancements in the blue-ward wing of the line in addition to the bright core of the line which indicates coronal upflows. A strong upflow with a 
velocity of 50 to 150~km~s$^{-1}$ have been reported by \citet{2011Sci...331...55D}. To distinguish the upflow emission component from the bright core, \citet{2009ApJ...701L...1D} and \citet{2011ApJ...738...18T} have used the asymmetry measurement in emission line profiles. To quantify the asymmetry, they subtracted red wing from blue wing (B-R) of the spectral lines. They show that the dominant primary emission component is superimposed with a faint secondary component which causes asymmetry in the line profiles. These quasi-periodic upflows create a profound asymmetry in the blue wing of the spectral line. The secondary component also enhances the line intensity and line width and cause a change in doppler shift periodically. 
\citet{2012ApJ...759..144T} found that footpoints of active region loops show oscillations with period around 10 min. They also found that all the line parameters (line intensity, Doppler shift, line width, and profile 
asymmetry) vary coherently and show apparent blueshifts and blue-ward asymmetry in the line profiles. 
They proposed that these oscillations are due to quasi-periodic upflows which supply hot plasma and energy to the corona. 
Quasi-periodic upflows have also been reported in coronal loops using spectroscopic observations by many authors \citep{2011ApJ...730...37U,2011ApJ...732...84M,2012ApJ...760L...5B,2012ASPC..455..361S,2013ApJ...779....1T}.
\citet{2012ApJ...754L...4T} have pointed out that upflows are strong at the loop footpoints and their strength decreases with height.

These recent observations challenge the well established explanation of PDs as slow magnetoacoustic waves. 
But, \citet{2010ApJ...724L.194V} have shown that due to the in-phase behaviour of velocity and density perturbations, 
upward propagating slow waves generally have the tendency to enhance the blue wing of the emission line.  
\citet{2012ASPC..456...91W} have used a different method (which includes photon noise) to measure the velocity of secondary emission component due to flows and
 showed that it was overestimated due to the saturation effects. They argued that the flow interpretation of the observed PDs is less favorable compared to the wave interpretation.
\citet{2011ApJ...737L..43N} found blueward asymmetry in the line profiles at the base of an active region but with increasing height along the loop, line profiles become symmetric. 
At higher locations intensity disturbances are in phase with the velocity which favours upward propagating slow-mode waves scenario. 

It is believed now that both the waves and flows might actually coexist close to the foot points of the loops 
and imaging observations alone are insufficient to distinguish them. One needs to perform a detailed analysis 
of all the line parameters using spectroscopic data which we attempt to do in this article.

\section{Observations And Data Reduction}
The dataset used in the present study is obtained from the observations of an active region AR11465 on 2012 April 26, by the EUV imaging spectrometer (EIS) 
onboard HINODE \citep{2007SoPh..243...19C,2007SoPh..243....3K} and the Atmospheric Imaging Assembly (AIA) \citep{2012SoPh..275...17L}, onboard Solar Dynamic Observatory (SDO). 
The details of these observations and the procedures employed to prepare the data for analysis, are given in this section.
\subsection{EIS}
\begin{figure}
\centering
\includegraphics[width=.5\textwidth,angle=90]{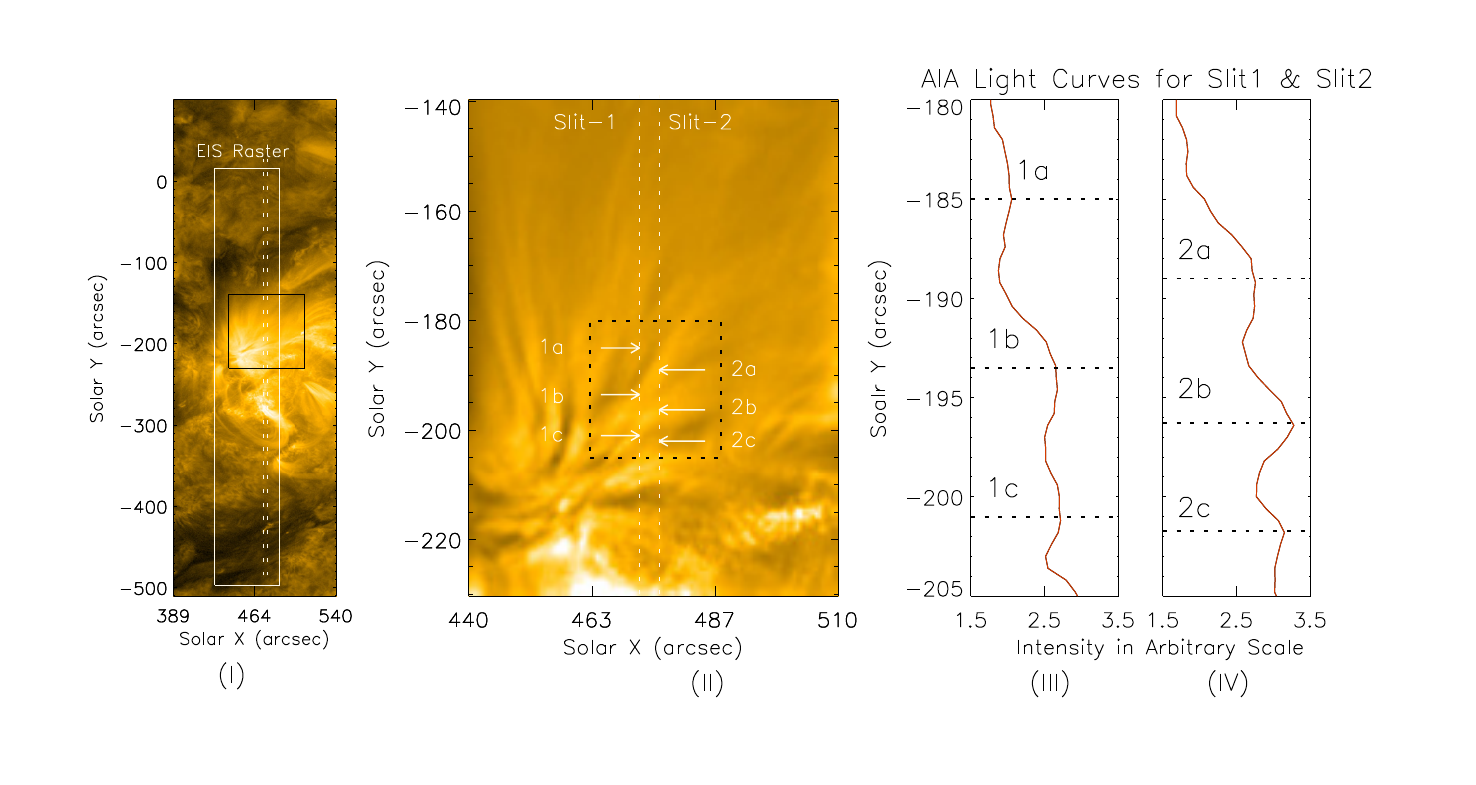}

\caption{(I) AIA 171~\r{A} image displaying fan-like loop structures
at an active region boundary. The white rectangular box marks the region covered by the EIS raster.
Two vertical dashed lines represent the positions of two EIS slits used for sit-and-stare observations. 
The small rectangular black box shows our Region Of Interest (ROI). 
(II) Zoomed-in view of the ROI showing the two slit positions and 6 analysis locations (1a-1c and 2a-2c). 
(III) \& (IV) Average 171~\r{A} AIA intensity profiles along slits 1 \& 2 respectively, showing the identification of loop crossing points (LCPs).}
\label{ms2290fig1}
\end{figure}

The EIS observation was taken in two modes. The first one is a raster scan obtained with the 2$''$ slit from 13:02 UT to 13:33 UT covering the active region via 30 raster steps. 
The X and Y pixel scales are $2''$ and $1''$ respectively. The Field Of View (FOV) covered by the raster is $60'' \times 512''$, shown as white box in Figure \ref{ms2290fig1}~(I). 
The second mode is a sit-and-stare observation obtained with the 2$''$ slit placed at two positions (as shown in white dashed line in Fig.\ref{ms2290fig1}~(I)) between 13.35 UT and 14.35 UT. 
The first sit and stare observation (slit-1) is obtained from 13:35 UT to 14:05 UT and the second one (slit-2) is from 14:05 UT to 14:35 UT. 
The cadence of these observations is 47 s and the total duration of each  set is 31 min. 
We performed the standard data preprocessing for all the EIS data with $eis\_prep$.pro (available as a part of the SolarSoft package) which includes dark current subtraction, cosmic ray removal, missing or saturated pixel flagging etc. 
It also corrects for the slit tilt and orbital variation. 

\subsection{AIA}
We have used the corresponding SDO/AIA data in the 171~\r{A} and 193~\r{A} channels taken from 13:00 UT to 15:00 UT which covers the EIS observation period. The cadence is 12 s. The level 1.0 data have been reduced to level 1.5 using the $aia\_prep.pro$ which makes the necessary instrumental corrections. The final pixel scale is $\approx$0.6$^{\prime\prime}$ in both  X and Y directions. Data from the AIA 193~\r{A} channel is used for coalighnment between the two instruments. The coalighnment has been achieved by cross-correlating the EIS 195~\r{A} raster scan with the corresponding AIA 193~\r{A} image. The final positions of the EIS sit-and-stare slits after correcting for the obtained offsets, are shown as vertical dashed lines in Figure~\ref{ms2290fig1}.

Figure~\ref{ms2290fig1} clearly shows the EIS slits positioned over the fan-like loop structures. We identify 6 locations, three (1a-1c) over the first slit and three (2a-2c) over the second slit, where the fan loops are found to cross the EIS slits. These locations, referred as Loop Crossing Points (LCPs), are identified from the peaks in time averaged AIA 171~\r{A} intensity profiles along the respective slits (see panels III and IV of Figure~\ref{ms2290fig1}).

\section{Data analysis and Results}
\begin{figure}
\centering
\includegraphics[width=1.\textwidth,angle=90]{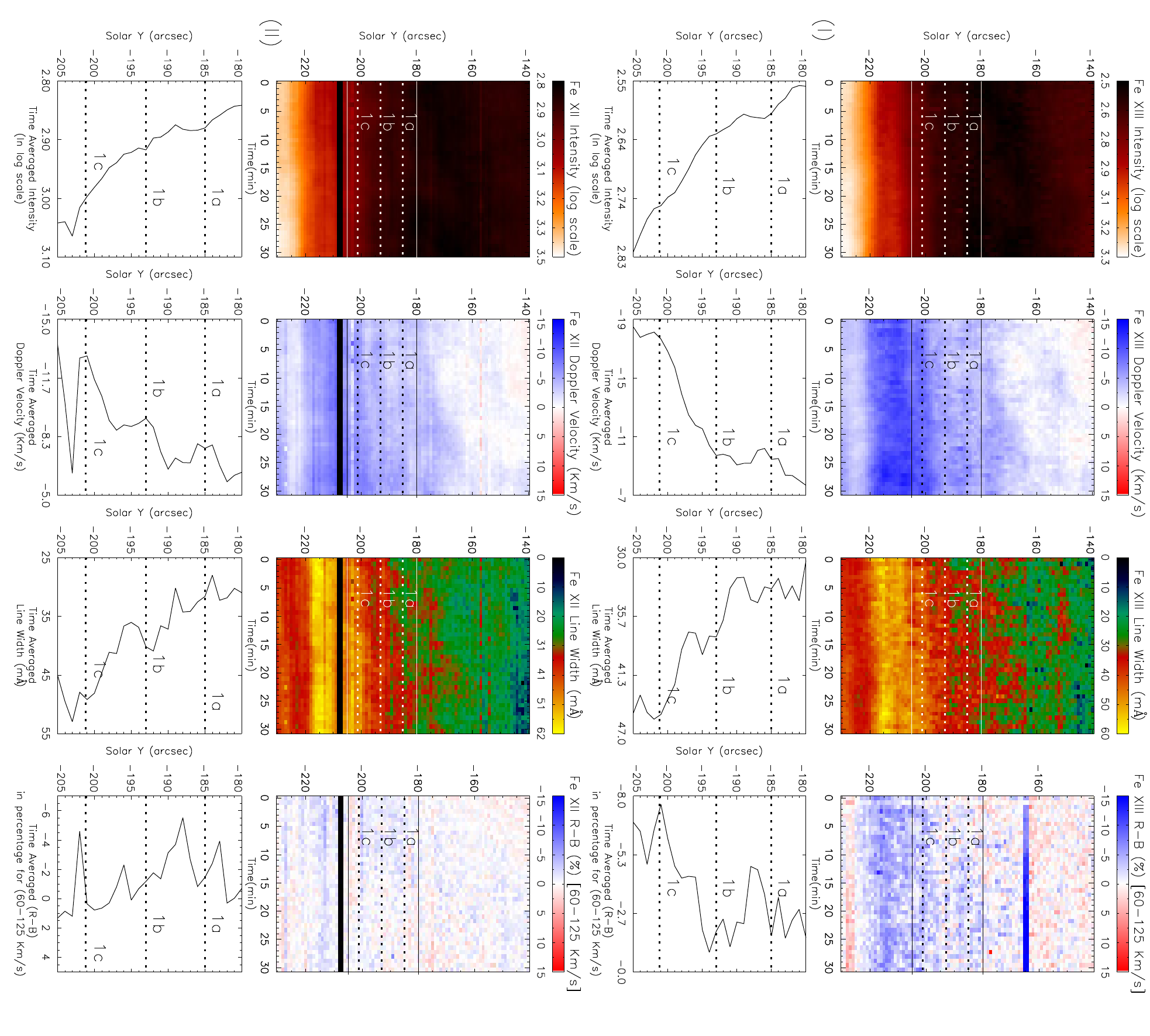}
\caption{(I) Top panels (from left to right) show the 
temporal evolution of the peak intensity, Doppler shift, line width, and R-B asymmetry, in 
the Fe \textsc{xiii} 202.04~\r{A} line for slit-1. Bottom panels show the time averaged values in these parameters for the region bounded by the solid lines marked in top panels. LCPs corresponding to slit-1 are marked in all these figures. (II) Same as the above for the Fe \textsc{xii} 195.12~\r{A} line. The black horizontal band near $Y \approx -207^{\prime\prime}$ correspond to missing data due to bad pixels.}
\label{ms2290fig2}
\end{figure}

Data from Fe \textsc{xii} 195.12~\r{A} and Fe \textsc{xiii} 202.04~\r{A} lines are used in the present analysis. The Fe \textsc{xii} 195.12~\r{A} line is self-blended with another Fe \textsc{xii} line at 195.18~\r{A} \citep{2009A&A...495..587Y}. So we have used a double Gaussian fitting for this line to derive all the spectral line parameters {\it i.e.} intensity, Doppler velocity, and line width using $eis\_auto$$\_fit.pro$. While fitting the line profile with two Gaussians, the separation between the lines was fixed at 0.06~\r{A} and the line widths were restricted to be same. Equal widths are expected since both the lines are from the same ion. The spectral parameters for the Fe \textsc{xiii} 202.04~\r{A} line are derived from a single Gaussian fitting. We also estimated the ``Red minus Blue'' (R-B) asymmetry in the line profiles by following a method similar to that described in \citet{2011ApJ...738...18T}. The exact procedure for obtaining the R-B asymmetry involves the following steps: (i) First we use spline interpolation to increase the line profile sampling by a factor of 10 times more than the original. (ii) Then we select the peak intensity position as the line centroid (following RB$_p$ method in the reference). (iii) Two narrow spectral windows are then selected at a chosen distance on both sides of the centroid {\it i.e.} at the red and at the blue wing. We have selected these spectral windows to be at wavelength positions corresponding to 60-125 km~s$^{-1}$. (iv) The total intensity in this window at the blue wing is then subtracted from that at the red wing to obtain the R-B value. For the 195.12~\r{A} line, the corresponding contribution from the 195.18~\r{A} line in the red wing is also subtracted. The obtained R-B values are normalised with the peak intensity to give (\%) asymmetry in the line profile. 

The constructed time-distance maps of line intensity, Doppler shift, line width, and R-B asymmetry, corresponding to the slit-1, for the EIS Fe \textsc{xii} 195.12~\r{A} and Fe \textsc{xiiI} 202.04~\r{A} lines are shown in Figure~\ref{ms2290fig2}. The bottom panels, for each spectral line, in this figure show the time-averaged variation in these parameters for the region bounded by the solid lines in the top panels. The LCPs corresponding to slit-1 are also marked in this figure. An interesting feature to note is the corresponding increase in Doppler shift and line width at the LCPs (\citet{2011ApJ...738...18T} found similar results). The R-B asymmetries for both the lines show enhancements in blue wings leading to a negative R-B value. 

\subsubsection{Coherence in Line Parameters}

\begin{figure}
\centering
\includegraphics[width=0.5\textwidth,angle=90,natwidth=1024,natheight=1024]{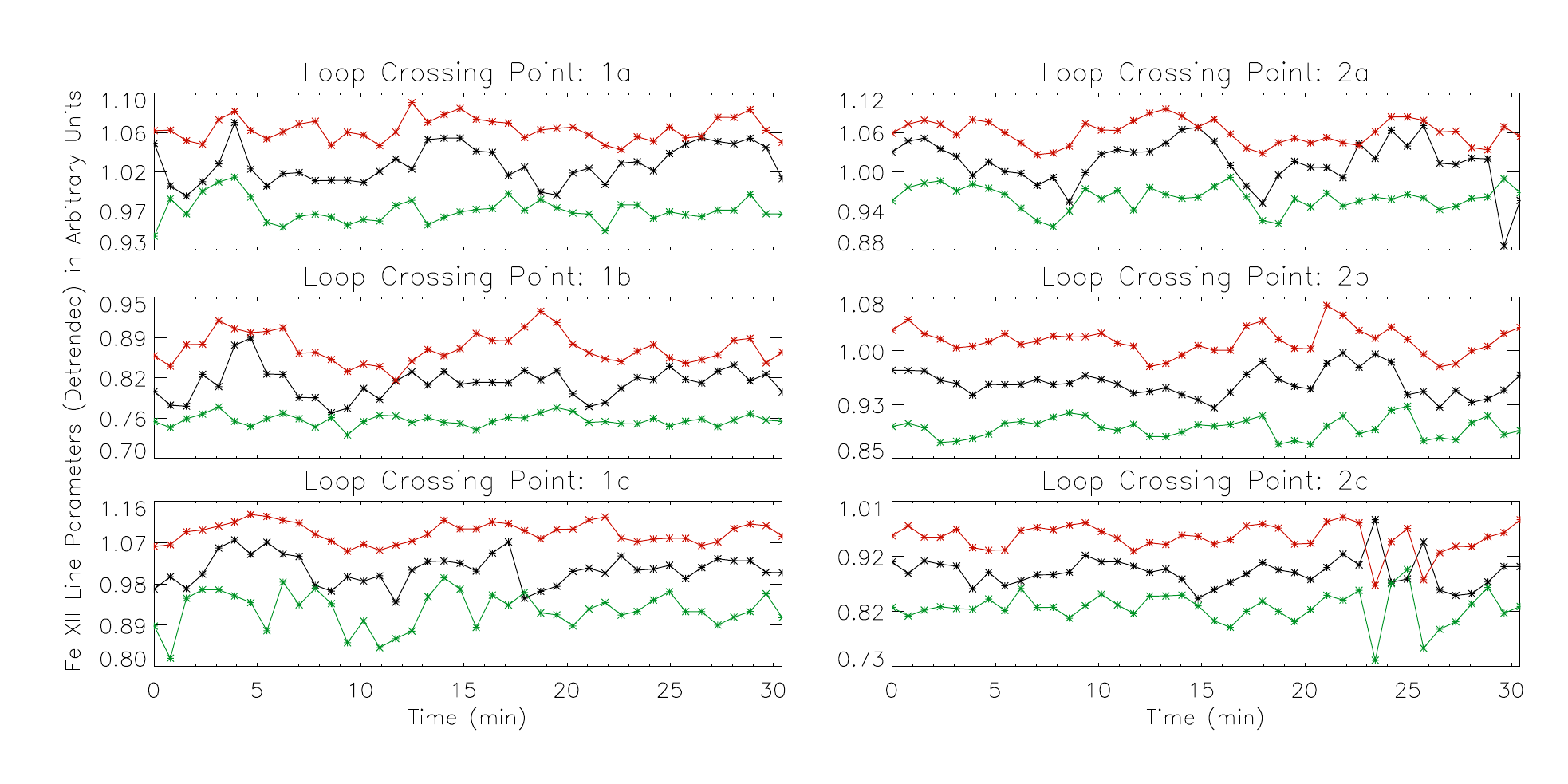}
\caption{ Variation in intensity, Doppler shift, and width of the Fe~\textsc{xii} 195~\AA\ line for all the LCPs is shown in red, black, and green curves respectively, in each panel.}
\label{ms2290fig3}
\end{figure}

 The temporal evolution of the EIS 195~\r{A} and 202.04~\r{A} lines at LCPs indicate oscillations in all the line parameters. We show the variation in intensity, Doppler shift, and line width, for all the LCPs (for 195.12~\r{A} line) in Figure~\ref{ms2290fig3}. All these light curves are trend subtracted by a 15 point running average to filter long period variations. To explore the possible coherency in oscillations and to quantify it, we computed cross-correlation coefficients between different line parameters for both the lines and listed them in Table \ref{cross}. Carefully looking at the table, we notice that the correlation values between intensity and velocity, are relatively high and persistent in all the LCPs for both the lines.

 \begin{table}[h]
\begin{center}
\caption{ Cross-correlation Coefficient values}  

\label{cross}
\begin{tabular}{lcccc r@{   }l c} 
  \hline
  & \multicolumn{1}{c}{$Position$ }& \multicolumn{1}{c}{$Intensity-$} & \multicolumn{1}{c}{$Doppl.Velocity-$} & \multicolumn{1}{c}{$Line Width-$} \\
&$        $ & $Doppl.Velocity$ & $Line Width$ & $Intensity$\\

& $ $ & $ 195.12 $ $  $   $  $ $ 202.04$ & $ 195.12 $ $  $   $  $ $ 202.04$ & $ 195.12 $ $  $   $  $ $ 202.04$ \\

     \hline
     &  1a & 0.51  $ $ $ $   0.62 & 0.18  $ $ $ $   0.64 & 0.36   $ $ $ $   0.55\\
     &  1b & 0.43  $ $ $ $   0.64 & 0.05  $ $ $ $   0.63 & 0.46   $ $ $ $   0.50\\
     &  1c & 0.55  $ $ $ $   0.15 & 0.24  $ $ $ $   0.52 & 0.56   $ $ $ $   0.24\\
     &  2a & 0.56  $ $ $ $   0.64 & 0.48  $ $ $ $   0.56 & 0.70   $ $ $ $   0.55\\
     &  2b & 0.53  $ $ $ $   0.57 & 0.20  $ $ $ $   0.30 & 0.60   $ $ $ $   0.68\\
     &  2c & 0.76  $ $ $ $   0.76 & 0.24  $ $ $ $   0.56 & 0.47   $ $ $ $   0.62\\
  \hline
\end{tabular}
\end{center}
\end{table}
%

\subsubsection{Periodicity of Oscillations} 

To measure the periodicity in the observed oscillations, we performed wavelet analysis on all the line parameters at the LCPs. Before performing the analysis, a 3-point running average has been considered to improve the signal-to-noise.

\begin{figure}
\centering
\includegraphics[width=0.28\textwidth,angle=90]{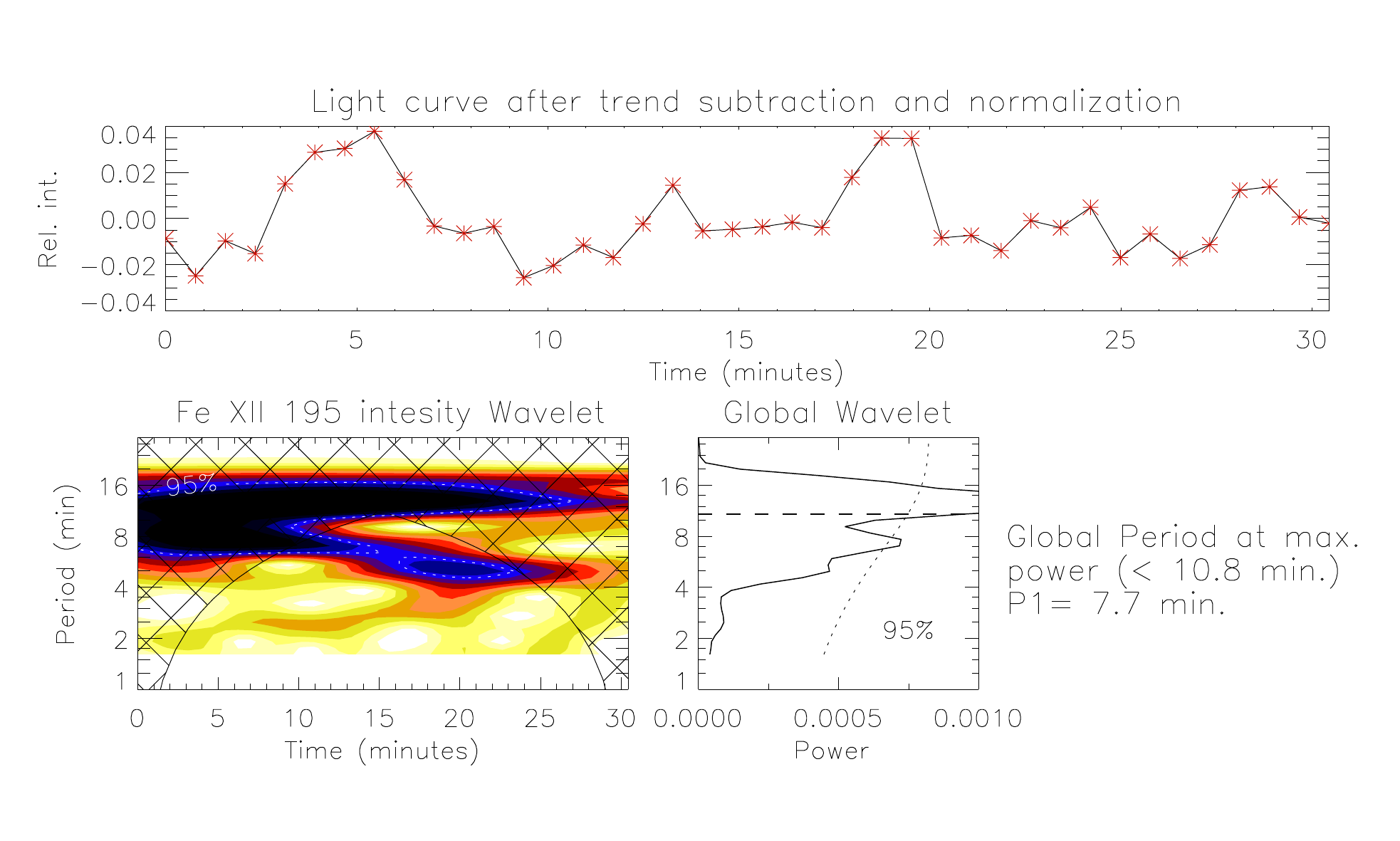}\includegraphics[width=0.28\textwidth,angle=90]{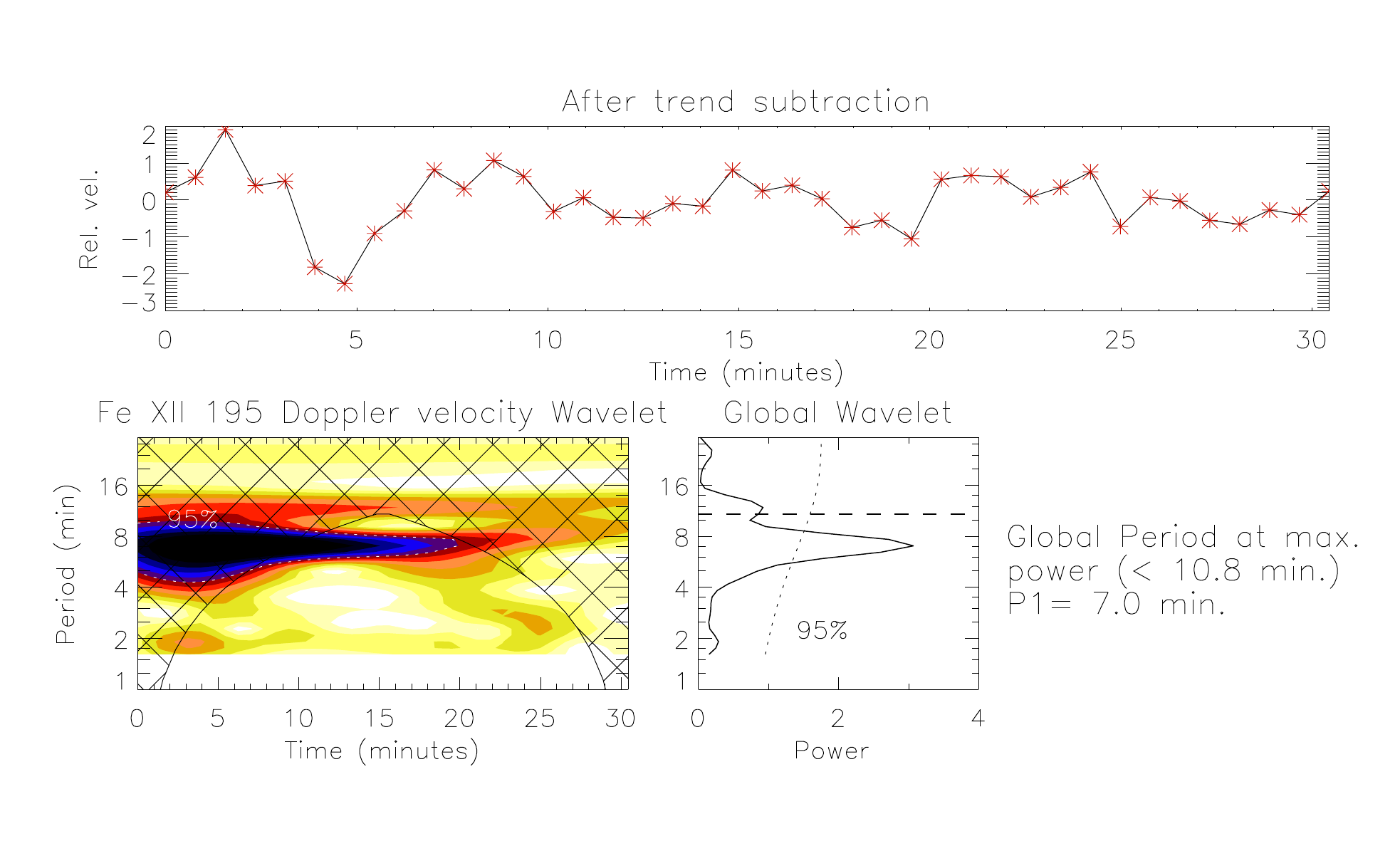}
\includegraphics[width=0.28\textwidth,angle=90]{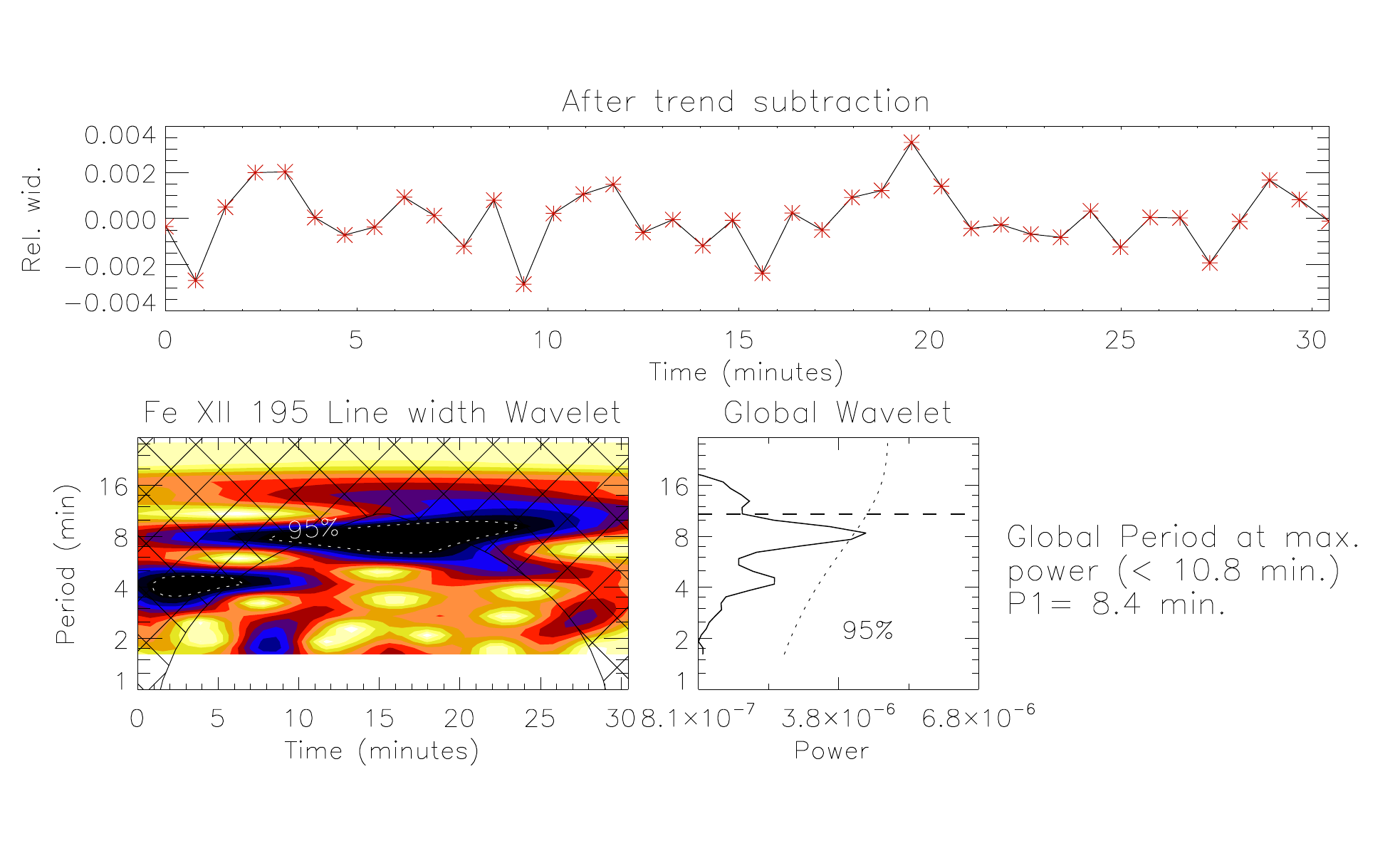}\includegraphics[width=0.28\textwidth,angle=90]{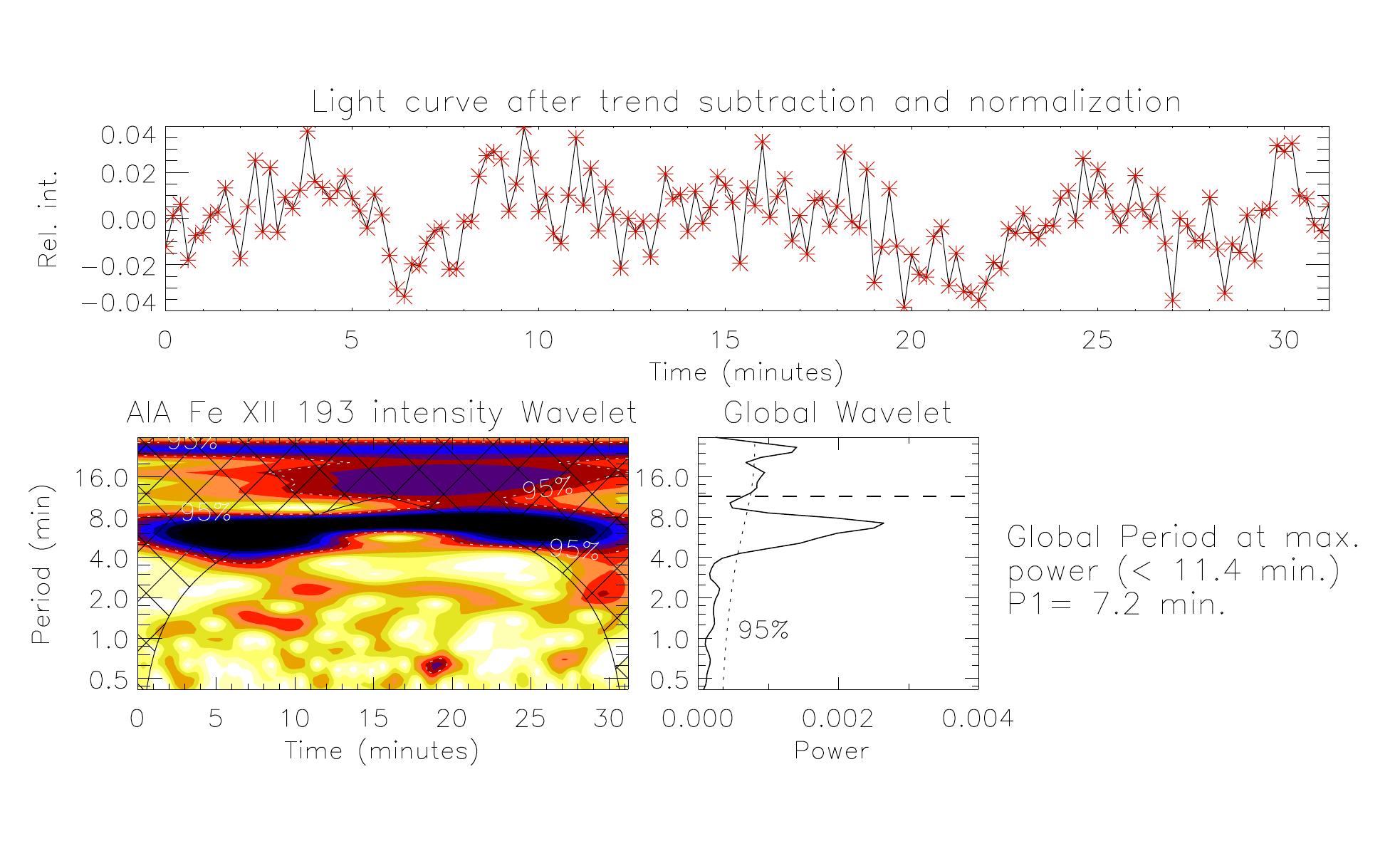}
\caption{Results of wavelet analysis for the variations in intensity (top left), Doppler velocity (top right), and width (bottom left) of EIS Fe \textsc{xii} 195~\r{A} line and AIA 193~\r{A} intensity (bottom right), corresponding to the LCP 1b. Top panel of each plot show the trend-subtracted light curve. Bottom-left panel shows the wavelet phase plot and the bottom-right panel shows the global power plot. The contours in the wavelet phase plot and the dotted curve in the global wavelet plot represent  95\% significance level for white noise \citep{1998BAMS...79...61T}. The periods with the maximum significant power are listed adjacent to the global wavelet plot.}
\label{ms2290fig4567}
\end{figure}

 The sample wavelet plots corresponding to the LCP 1b are shown in Figure~\ref{ms2290fig4567}. Wavelet results for the corresponding AIA intensity are also shown in this figure (bottom-right plot). In all these plots, the upper panel shows the trend-subtracted light curve, the bottom-left panel shows the wavelet power spectrum and the bottom-right panel shows the global wavelet power which is nothing but the wavelet 
power at each frequency scale averaged over time. The contours in the wavelet plot and the dotted line in the global wavelet plot represent the 95\% significance level calculated for white noise \citep{1998BAMS...79...61T}. The cross-hatched region in the wavelet plot indicates the cone of influence (COI) where edge effects come into play. In the global wavelet plot, the horizontal dashed line marks the maximum measurable period (due to COI), which is $\approx$11 min for our 31 min observation. The periods with the maximum significant power are listed adjacent to the global wavelet plot. These plots 
suggest a  periodicity of 7.7 min in EIS 195.12~\r{A} intensity, 7.0 min in Doppler velocity, 8.4 min in line width, and  7.2 min in AIA 193~\r{A} intensity at LCP 1b. The results for other LCPs (in other EIS lines and AIA channels) are listed in Table~\ref{period} which are in the range of 3 to 10 min.
  \begin{table}[h]
 \begin{center}
 \caption{Periodicities obtained from the wavelet analysis (in min)}  
 
 \label{period}
\begin{tabular}{lcccccc}
   \hline
   & \multicolumn{1}{c}{$Position$ } & \multicolumn{1}{c}{$ EIS $} & \multicolumn{1}{c}{$ EIS $} & \multicolumn{1}{c}{$ EIS $} & \multicolumn{1}{c}{$   AIA  $} \\

  & \multicolumn{1}{c}{$ $} & \multicolumn{1}{c}{$Intensity$} & \multicolumn{1}{c}{$Doppler Velocity$} & \multicolumn{1}{c}{$Line Width$} & \multicolumn{1}{c}{$  Intensity$} \\

& $ $ & $ 195.12 $ $  $   $  $ $ 202.04$ & $ 195.12 $ $  $   $  $ $ 202.04$ & $ 195.12 $ $  $   $  $ $ 202.04$ & $ 171$ $  $   $  $ $ 193$ \\
   \hline
      &  1a & 7.7   $ $ $ $   5.0& 7.7   $ $ $ $   4.6& 5.0   $ $ $ $   5.4& 7.9   $ $ $ $   7.2\\
      &  1b & 7.7   $ $ $ $   7.0& 7.0   $ $ $ $   7.7& 8.4   $ $ $ $   7.7& 7.2   $ $ $ $   7.2\\
      &  1c & 8.4   $ $ $ $   8.4& 5.9   $ $ $ $   4.2& 5.4   $ $ $ $   5.4& 7.9   $ $ $ $   6.1\\
      &  2a & 5.9   $ $ $ $   7.0& 7.7   $ $ $ $   7.0& 5.9   $ $ $ $   6.5& 6.1   $ $ $ $   6.6\\
      &  2b & 7.0   $ $ $ $   9.1& 4.6   $ $ $ $   9.1& 5.0   $ $ $ $   9.0& 6.1   $ $ $ $   6.1\\
      &  2c & 3.8   $ $ $ $   6.5& 7.0   $ $ $ $   5.9& 3.5   $ $ $ $   6.5& 5.1   $ $ $ $   9.4\\
   \hline
 \end{tabular}
 \end{center}
 \end{table}

\subsection{Image Analysis}

\subsubsection{Powermaps}

We have created a movie (Movie~1, available online) of the ROI from the AIA 171~\r{A} image sequence. One can clearly see the presence of PDs travelling outward in the movie. However, it requires further analysis to figure out their relation with the oscillations observed at LCPs.
\begin{figure}[h]
\centering
\includegraphics[width=0.75\textwidth,angle=90]{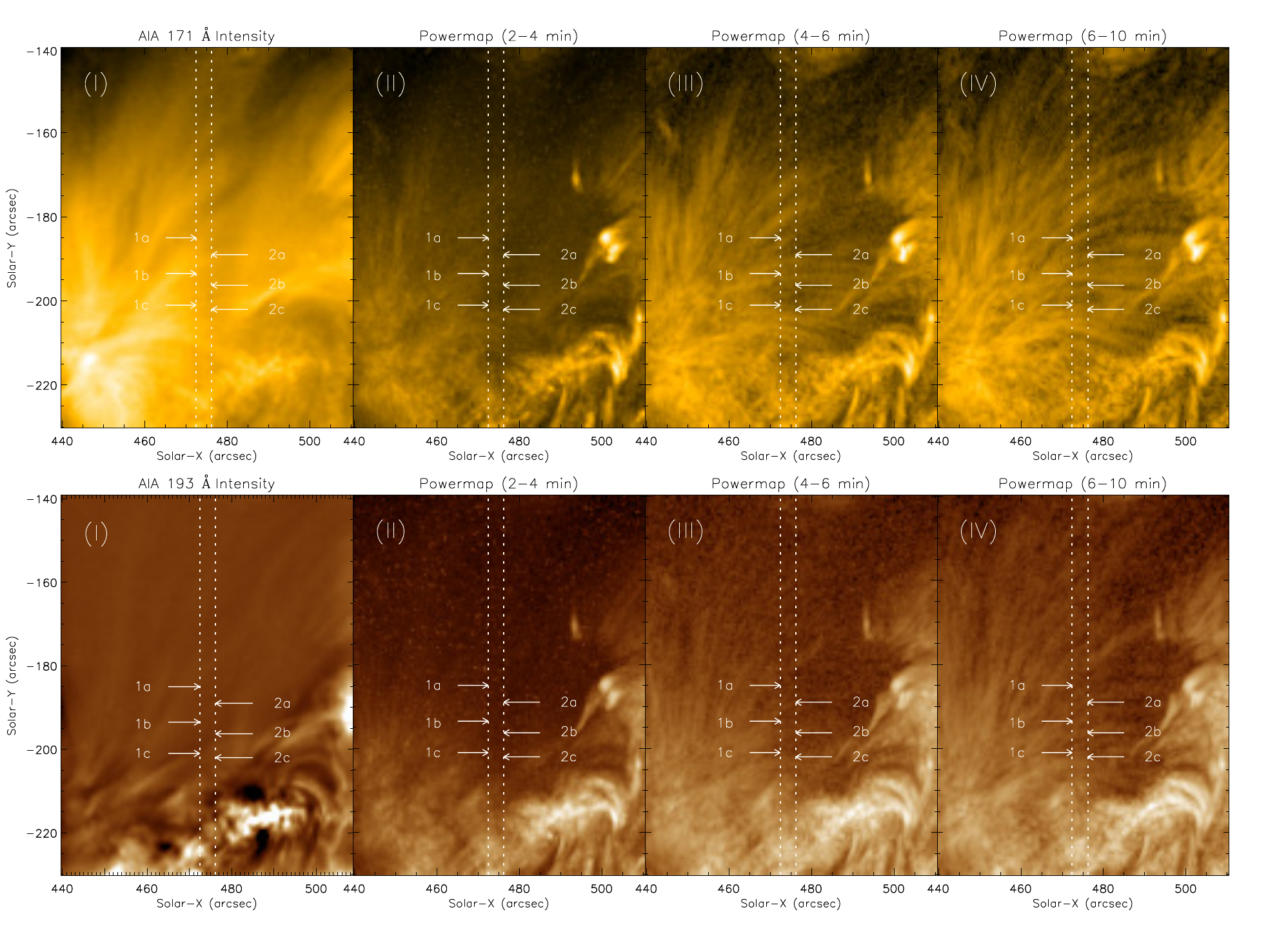}
\caption{AIA 171~\r{A} (top panel) and 193~\r{A} (bottom panel) intensity images for the ROI shown in Figure~\ref{ms2290fig1} and the corresponding powermaps in three period bands as indicated. The locations of the EIS slits and selected LCPs are also marked. The loops which are faintly visible in the intensity image of the 193~\r{A} channel are seen clearly in the respective powermaps.}
\label{ms2290fig8}
\end{figure}

 Using AIA images, we constructed powermaps of the selected region in three period bands, 2 -- 4 min, 4 -- 6 min, and 6 -- 10 min (see Figure~\ref{ms2290fig8}). To create these maps, we perform wavelet analysis on the light curve at each pixel within the selected region and obtain the power at all possible periods. Then the power in a selected period window is averaged to make a powermap corresponding to that period range. Loop-like structures are clearly visible in the powermaps and in the 6 -- 10 min period range which is observed at LCPs, these structures are visible up to the EIS slit positions. This might indicate that the PDs travelling along the loops, and the oscillations observed at the LCPs are related.

\subsubsection{Propagation Speeds}

\begin{figure}[h]
\centering
\includegraphics[width=0.75\textwidth,angle=90,natwidth=1048,natheight=500]{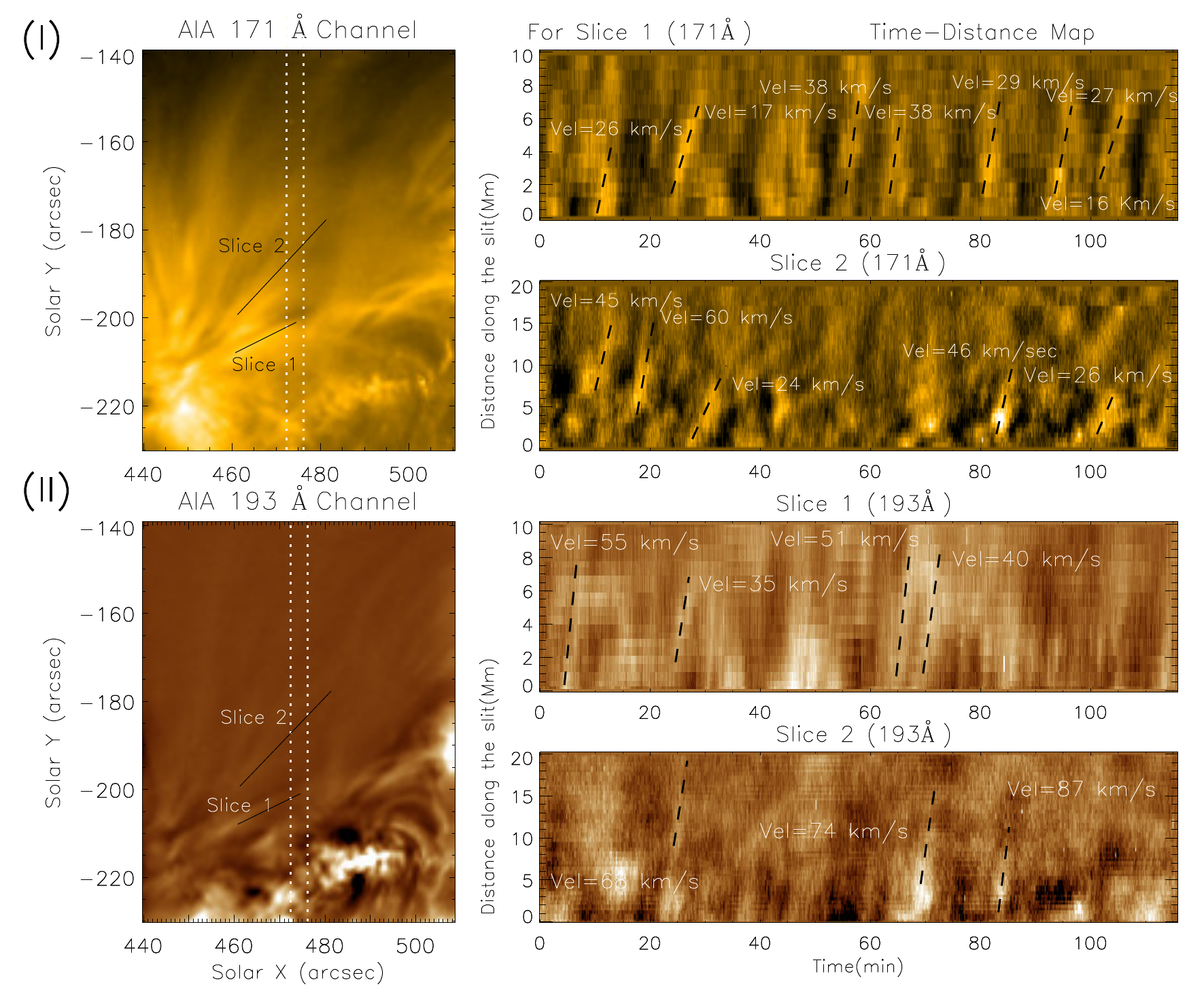}
\caption{(I) {\it Left}: AIA 171~\r{A} image showing the locations of the slices chosen for time-distance analysis. The vertical dashed lines mark EIS slit positions. {\it Right}: Enhanced time-distance maps constructed from slice 1 (top) and slice 2 (bottom). 
The inclined black dashed lines represent the slope of the individual ridges used in the propagation speed estimation. (II) Corresponding plots for AIA 193~\r{A} channel.}
\label{ms2290fig9}
\end{figure}

Propagation speed is one of the important parameters to understand the nature of PDs. To examine the typical speeds in this region we performed time-distance analysis. Two fan loops, crossing the EIS slits, were selected and artificial slices were made along those loops. Time-distance maps were then constructed for these slices using the AIA image sequence. The locations of the two slices and the corresponding time-distance maps, for the two AIA channels, are shown in Figure~\ref{ms2290fig9}. The time-distance maps were enhanced by subtracting longer trends at each spatial position. Bright ridges of varying intensity and inclination are visible in these maps. These ridges represent the observed PDs and their inclination gives the apparent propagation speed. The positions of the local maxima were identified along each ridge and fitted with a linear function to calculate the propagation speed. The obtained speeds for the individual ridges are marked in the figure. The values range from 17 -- 60 km~s$^{-1}$ (with errors less than 7 km~s$^{-1}$) for the 171~\r{A} channel and 37 -- 87 km~s$^{-1}$ (with errors less than 11 km~s$^{-1}$) for 193~\r{A} channel. The speed ratios, calculated from the common ridges appearing in the two AIA channels, range from 1.1 to 2.0. This shows the speeds of the PDs are temperature dependent. 
Note that these measured speeds are the apparent speeds in the plane of sky. Hence, the real speeds can be much higher.
\section{Summary and Conclusions}

We have studied the properties of PDs in an active region fan loop system using simultaneously observed imaging data from SDO/AIA and spectroscopic data from Hinode/EIS. At the 6 loop-crossing locations, identified from the AIA 171~\r{A} images, we observe oscillations in all the line parameters for the two EIS lines (Fe \textsc{xii} 195.12~\r{A} and Fe \textsc{xiii} 202.04~\r{A} ). The periodicity of these oscillations ranges from 3 to 10 min. The intensity oscillations obtained from AIA 193~\r{A} and 171~\r{A} also show a similar periodicity. AIA image sequence and the powermaps of the region clearly indicate that these oscillations are connected with the PDs propagating along the fan loops. The apparent propagation speeds of PDs were found to be different for the AIA 193 (T$\approx$1.3 MK) and AIA 171 (T$\approx$0.6 MK) channels. The average speed ratio is $\approx$ 1.5 compared to the theoretical value 1.47 expected for these two channels. 

We also performed correlation analysis to find coherency between different line parameters which reveals a relatively high correlation between intensity and Doppler shift at all the LCPs. It may be noted that in some cases the periods of different line parameters are not exactly the same (see Table~\ref{period}). The listed values in Table~\ref{period} are only the dominant periods which are in most cases accompanied by other peaks (although not significant). The existence of multiple periods often degrades the correlation value. It may be useful to filter the time series to improve this but the small amplitudes of oscillations combined with the noise make it difficult to achieve.

The temperature-dependent propagation speeds and relatively high correlation between intensity and Doppler shift which are expected for a propagating slow magneto-acoustic wave, favour the wave interpretation for the observed PDs. On the other hand, the observed PDs were not so regular, with changing inclinations and intensities from one ridge to the other as can be seen from the time-distance maps. The R-B asymmetry analysis also reveals a negative R-B value for both the EIS lines indicating a blue wing enhancement. These properties suggest the possible interpretation of the observed PDs as quasi-periodic upflows. In fact, some of the latter properties can actually be explained with a propagating slow wave scenario if one considers the line profile resulting from the superposition of a high-speed component on a nearly stationary background. This superposition leads to profile asymmetry producing a blue-wing enhancement and also results in coherent oscillations in all the line parameters \citep{2010ApJ...724L.194V,2008ApJ...678L..67H}.

Based on the observed properties, we suggest a possible co-existence of waves and flows causing the PDs found in this region. Here we want to emphasise that it is very difficult to distinguish these two effects from each other from the intensity images as well as from the spectroscopic observations as presented here. Recently \citet{2015SoPh..290..399D} made synthetic spectra to justify one scenario over the other and they found that most of the observables which can give us an indication for the same (between wave and flow scenario), are not measurable with the current instrument capabilities. Efforts shall be made to combine theoretical modelling with observations which will help us to decouple these two phenomena from each other.

\section{Acknowledgments}
Hinode is a Japanese mission developed and launched by ISAS/JAXA, with NAOJ as
domestic partner and NASA and STFC (UK) as international partners. It is operated
by these agencies in co-operation with ESA and NSC (Norway).
The AIA data used here is courtesy of the SDO (NASA) and
AIA consortium.


\bibliographystyle{raa.bst}
\bibliography{references}

\clearpage

\end{document}